# L'articolo di Ettore Majorana su "Il valore delle Leggi Statistiche nella Fisica e nelle Scienze Sociali" (Ettore Majorana's article on "The value of Statistical Laws in Physics and in Social Sciences")

by  Erasmo Recami

*(Università statale di Bergamo, Bergamo, Italy;* and
INFN-Sezione di Milano, Milan, Italy)

**PRESENTATION and Abstact (in English):** The article "Sul valore delle Leggi Statistiche nella Fisica e nelle Scienze Sociali" was written by Ettore Majorana, in a partially educational way, for a journal of Sociology; but he gave up, afterwards, publishing it  (and, even more, threw it away). It appeared posthumous, thanks to Giovanni Gentile Jr. (the inventor of "parastatistics", and a great friend of Majorana) in the journal "Scientia", vol.36,  pp.58-66, issue of Feb.-Mar., 1942. It has not been re-published afterwards, in Italian, till the beginning of 2006, when we made some abridgements of it known by various Italian newspapers and by the journal "Fisica in Medicina". We don't know when was it written: perhaps in 1930, since it refers to standard Quantum Mechanics, without hints to the discussions risen in the thirthies. However, the central theme of this writing was still alive in Majorana's mind in 1934: in fact, on July 27, 1934, he will write to G.Gentile Jr. to expect that <<soon it will be generally understood that science ceased to be a justification for the vulgar materialism>>.  Here, we present in the First Part a suitable *reduction*, edited by us, of such an article; and, in the Second Part, a complete transcription of it. Since the paper which appeared in "Scientia" contains some errors in the interpretation of Majorana's handwriting, the present versions have been very slightly "corrected" by us.  The complete text has been translated once into French, and twice into English: of the latter, it is known to us R.Mantegna's, published at the end of 2006 [in the book "E.Majorana: Scientific Papers", ed. by  F.Bassani et al. (SIF and Springer; Bologna and Berlin)].  The interested reader can find all the known biographical documents --apart from the ones discovered during the last two years-- in the book by E.Recami, "Il Caso Majorana: Epistolario, Testimonianze, Documenti" (Mondadori, Milan, 1987 and 1991; Di Renzo Editore, Rome, 2000 and 2002); and in the e-prints arXiv:physics/9810023v4 [physics.hist-ph]; arXiv:0708.2855v1 [physics.hist-ph]; arXiv:0709.1183 [physics.hist-ph].  As the Abstract, let us report, from Majorana's summary: <<The *deterministic* conception of nature contains into itself a real cause of weakness, due to the unrectifiable contradiction it meets with the most certain data coming from our own consciousness. G.Sorel attempted at settling such a disagreement by distinguishing between *artificial nature* and *natural nature* (the latter being a-causal), but in such a way he denied the unity of science. Furthermore, the formal analogy between the statistical laws of Physics and those of Social Sciences fostered the opinion that also the human issues were submitted to a rigid determinism. It is therefore important that the recent principles of Quantum Mechanics have led us to recognize (besides a certain lack in objectivity of the phenomena description) the *statistical* character also of the basic laws of elementary processes. Such a result has made substantial the analogy between physics and social sciences, and indicated an identity in value and method between them.>>

Let us add that one interesting observation in this early paper by Majorana (written possibly at the end of the twenties, as we were saying) is that by easy experimental set-ups one can prepare a complex and showy chain of phenomena *started* by the accidental disintegration of a single radioactive atom: From the strictly scientific point of view nothing forbids us considering as plausible that some human events too can be originated by an equally simple, invisible and unpredictable occurrence.

*  *  *

***PRESENTAZIONE e Riassunto in Italiano:** Il presente articolo fu scritto da Ettore Majorana, in maniera parzialmente didascalica, per una rivista di sociologia; rinunciando poi a pubblicarlo (ed, anzi, cestinandolo). Esso ha visto la luce postumo, per interessamento di Giovanni Gentile jr., (inventore delle "parastatistiche" e grande amico di Ettore, sulla rivista* **Scientia**, *vol.36, fascicolo del Febbraio-Marzo del 1942, pp.58-66. Dopo di allora, non è stato più ripubblicato in lingua italiana fino agli inizi del 2006, quando ne abbiamo reso note varie riduzioni su differenti quotidiani italiani e sulla rivista "Fisica in Medicina". Non si sa quando fu scritto: forse nel 1930, dato che si fa riferimento alla meccanica quantistica standard, senza accenni alle critiche sorte negli anni trenta. Però il tema centrale di questo scritto era ancora vivo nell'animo del Nostro nel 1934: infatti, il 27.07.34 (su carta listata a lutto, dato che quell'anno era mancato suo padre), il Majorana scriverà a Giovannino Gentile di attendersi che <<presto sarà generalmente compreso che la scienza ha cessato di essere una giustificazione per il volgare materialismo>>. Nella Prima Parte di questo scritto presentiamo una opportuna riduzione di tale articolo, a nostra cura; mentre nella Seconda parte ne riportiamo la versione integrale. Dato che il testo apparso su* **Scientia** *contiene*

*alcuni evidenti errori, commessi nell'interpretazione della calligrafia di Majorana, le presenti versioni sono state pure leggermente "corrette".*
*Il testo completo è stato tradotto in francese e, due volte, in inglese: di queste ultime, conosciamo la traduzione di R.Mantenga, apparsa alla fine del 2006 [nel libro "E.Majorana: Scientific Papers", a cura di F.Bassani et al. (SIF e Springer; Bologna e Berlino)]. Il lettore interessato potrà trovare tutti i documenti biografici noti –a parte quelli rinvenuti nell'ultimo anno- nel volume di E.Recami su "Il Caso Majorana: Epistolario, Testimonianze, Documenti" (Mondadori, Milano, 1987 e 1991; Di Renzo Editore, Roma, 2000 e 2002).; and in the e-prints arXiv:physics/9810023v4 [physics.hist-ph]; arXiv:0708.2855v1 [physics.hist-ph]; arXiv:0709.1183 [physics.hist-ph]. Quale Riassunto, riportiamo anzitutto il sommario scritto da Majorana: <<La concezione deterministica della natura racchiude in sè una reale causa di debolezza nell'irrimediabile contraddizione che essa incontra con i dati più certi della nostra stessa coscienza. G.Sorel tentò di comporre questo dissidio con la distinzione tra natura artificiale e natura naturale (quest'ultima acausale), ma negò così l'unità della scienza. D'altra parte l'analogia formale tra le leggi statistiche della Fisica e quelle delle Scienze Sociali accreditò l'opinione che anche i fatti umani sottostassero a un rigido determinismo. E' importante, quindi, che i recenti principii della Meccanica Quantistica abbiano portato a riconoscere (oltre ad una certa assenza di oggettività nella descrizione dei fenomeni) il carattere statistico anche delle leggi ultime dei processi elementari. Questa conclusione ha reso sostanziale l'analogia tra fisica e scienze sociali, tra le quali è risultata un'identità di valore e di metodo.>> Aggiungiamo che una interessante osservazione di questo precoce articolo del Majorana (scritto forse nel 1930, come si diceva) è che comuni artifici di laboratorio bastano per preparare una catena comunque complessa e vistosa di fenomeni che sia comandata dalla disintegrazione accidentale di un solo atomo radioattivo; e che non vi è nulla dal punto di vista strettamente scientifico che impedisca di considerare come plausibile che all'origine di avvenimenti umani possa trovarsi un fatto vitale egualmente semplice, invisibile e imprevedibile.*

# PARTE PRIMA

In questa Prima Parte pubblichiamo una versione opportunamente ridotta dell'articolo di Ettore Majorana. La riduzione (e semplificazione) è stata operata avendo cura di mettere in evidenza tutte le parti sostanziali e più innovative dello scritto qui considerato:

## ETTORE MAJORANA:
### Il valore delle Leggi Statistiche nella Fisica e nelle Scienze Sociali

*Riassunto dell'Autore*: La concezione *deterministica* della natura racchiude in sè una reale causa di debolezza nell'irrimediabile contraddizione che essa incontra con i dati più certi della nostra stessa coscienza. G. Sorel tentò di comporre questo dissidio con la distinzione tra *natura artificiale* e *natura naturale* (quest'ultima acausale), ma negò cosi l'unità della scienza. D'altra parte l'analogia formale tra le leggi statistiche della Fisica e quelle delle Scienze Sociali accreditò l'opinione che anche i fatti umani sottostassero a un rigido determinismo. E' importante, quindi, che i recenti principii della Meccanica Quantistica abbiano portato a riconoscere (oltre ad una certa assenza di oggettività nella descrizione dei fenomeni) il carattere *statistico* anche delle leggi ultime dei processi elementari. Questa conclusione ha reso sostanziale l'analogia tra fisica e scienze sociali, tra le quali è risultata un'identità di valore e di metodo.

*********

È noto che le leggi della meccanica, in modo particolare, sono apparse lungamente come il tipo insuperabile delle nostre conoscenze della natura, e si è anzi creduto da molti che a tal tipo, in ultima analisi, si sarebbero dovute ricondurre anche tutte le altre scienze. Valga ciò di giustificazione allo studio che intraprendiamo.

## 1. - LA CONCEZIONE DELLA NATURA SECONDO LA FISICA CLASSICA.

Il credito goduto dalla fisica deriva dalla scoperta delle cosiddette "leggi esatte", consistenti in formule relativamente semplici, che si rivelano di universale validità, sia che vengano applicate a nuovi ordini di fenomeni, sia che il progressivo affinamento dell'arte sperimentale le sottoponga a un controllo sempre più rigoroso. È a tutti noto che, secondo la meccanica classica, il movimento di un corpo materiale è *interamente determinato* dalle condizioni iniziali (posizione e velocità) in cui il corpo si trova, e dalle *forze* che agiscono su di esso... In un caso è stato possibile trovare l'espressione generale di queste forze: nel caso cioè che i corpi interagenti siano isolati e agiscano quindi reciprocamente solo *a distanza...*; una situazione di questo tipo la si incontra in presenza della gravitazione universale di Newton, la quale è tipicamente applicabile allo studio dei movimenti degli astri. Come è noto, tale legge è realmente sufficiente per prevedere in ogni aspetto e con esattezza meravigliosa tutto il complesso svolgimento del nostro sistema planetario. [Una sola minuta eccezione, riguardante lo spostamento secolare che subisce il perielio di Mercurio, costituisce una delle maggiori prove sperimentali della recente teoria della relatività generale].

Il successo sensazionale della meccanica applicata all'astronomia ha incoraggiato la supposizione che anche i fenomeni più complicati dell'esperienza comune debbano infine ricondursi a un meccanismo simile, e solo alquanto più generale, della legge di. gravitazione. Secondo tale modo di vedere, che ha dato luogo alla *concezione meccanicistica* della natura, tutto l'universo materiale si svolge obbedendo a una legge inflessibile, in modo che il suo stato in un certo istante è interamente determinato dallo stato in cui si trovava nell'istante precedente; segno che tutto il futuro può essere previsto con assoluta certezza purché lo stato attuale dell'universo sia interamente noto. Tale concezione pienamente deterministica della natura sembra avere avuto in seguito numerose conferme; gli sviluppi ulteriori della fisica classica sembrano vigorosamente confermare il punto essenziale, cioè la completa *causalità* fisica.

Non è contestabile che si debba proprio al determinismo il merito principale di aver reso possibile il grandioso sviluppo moderno della scienza, anche in campi lontanissimi dalla fisica. *Eppure il determinismo, che non lascia alcun posto alla libertà umana e obbliga a considerare come illusori, nel loro apparente* finalismo, *tutti i fenomeni della vita, racchiude una reale causa di debolezza: la contraddizione immediata e irrimediabile con i dati più certi della nostra coscienza.*

Come il suo effettivo e, secondo ogni verosimiglianza, definitivo superamento sia avvenuto proprio nella fisica in questi ultimi anni, diremo più avanti; sarà anzi nostro scopo l'illustrare il rinnovamento che il concetto tradizionale delle leggi statistiche deve subire in conseguenza del nuovo indirizzo seguito dalla fisica contemporanea. Ma per il momento vogliamo ancora attenerci alla concezione classica della fisica, essendo essa ancora la sola largamente conosciuta oltre la cerchia degli specialisti. Prima di chiudere questa parte introduttiva, crediamo opportuno ricordare che le critiche al determinismo si sono nel tempo via via moltiplicate..., invocando alcune volte un principio metafisico di G.B.Vico, e più spesso il principio pragmatista. Quest'ultimo –il principio di giudicare le dottrine scientifiche in base alla loro concreta utilità-- non giustifica in alcun modo, però, la pretesa di condannare l'ideale dell'unità della scienza, che si è rivelata più volte un efficace stimolo al progresso delle idee.

## 2. - IL SIGNIFICATO CLASSICO DELLE LEGGI STATISTICHE E LE STATISTICHE SOCIALI.

Per bene intendere il significato delle leggi statistiche secondo la Meccanica, riferiamoci alla struttura interna dei corpi gassosi, la quale è particolarmente semplice. Infatti, nei gas in condizioni ordinarie le singole molecole xhe li costituiscono si possono considerare come particolarmente indipendenti, e a distanze reciproche considerevoli rispetto alle loro ridottissime dimensioni… [*omissis*]… Vi è una intera branca della fisica, la termodinamica, i cui principii si possono ricondurre alle nozioni generali della meccanica statistica. Per quanto abbiamo fatto finora, si può così riassumere il significato delle leggi statistiche secondo la fisica classica: 1°) i fenomeni naturali obbediscono ad un determinismo assoluto; 2°) l'osservazione *ordinaria* non permette di riconoscere esattamente lo stato *interno* di un corpo, ma solo il suo stato macroscopico; 3°) stabilite delle ipotesi plausibili… il calcolo delle probabilità permette la previsione più o meno certa dei fenomeni futuri.

Possiamo ormai esaminare il rapporto che passa fra le leggi stabilite dalla meccanica classica e quelle regolarità empiriche che sono note con lo stesso nome in modo particolare nelle scienze sociali. L'analogia formale, infatti, non potrebbe essere più stretta. Quando si enuncia, ad esempio, una legge statistica su una certa popolazione, è chiaro che si rinuncia deliberatamente a indagare sulla biografia degli gli individui che compongono la società in esame; non altrimenti, allorché si definisce lo stato (macroscopico) di un gas semplicemente dalla pressione e dal volume, si rinuncia deliberatamente a investigare posizione e velocità di tutte le singole molecole costituenti.

Ammesse le ragioni che fanno credere all'esistenza di una reale analogia fra le leggi statistiche fisiche e sociali, si potrebbe essere indotti a ritenere che, come le prime presuppongono logicamente un rigido determinismo, così le ultime possano essere ritenute da parte loro la prova che il determinismo governa anche i fatti umani; argomento avvalorato dalla tendenza a vedere nella causalità della fisica classica un modello di valore universale.

Sarebbe qui fuor di luogo riprendere discussioni antiche e mai concluse, ma va accolto con viva attenzione l'annunzio che negli ultimissimi anni la fisica è stata costretta ad abbandonare il suo indirizzo tradizionale rigettando, in maniera verosimilmente definitiva, il determinismo assoluto della meccanica classica.

### 3. - LE NUOVE CONCEZIONI DELLA FISICA.

È impossibile esporre con qualche compiutezza in poche righe lo schema matematico e il contenuto sperimentale della meccanica quantistica: ci limiteremo pertanto a qualche accenno. Vi sono dei fatti sperimentali noti da gran tempo (fenomeni di interferenza) che depongono irrefutabilmente a favore della teoria *ondulatoria* della luce; altri fatti scoperti da recente (effetto Compton) suggeriscono, *al contrario*, non meno decisivamente l'opposta teoria *corpuscolare*. Tutti i tentativi di comporre la contraddizione nel quadro della fisica classica sono rimasti infruttuosi. Senonché di tali fatti inesplicabili, e di molti altri, si è trovata da pochi anni una spiegazione unica e meravigliosamente semplice: quella contenuta nei principii della *meccanica quantistica*.

Gli aspetti caratteristici della meccanica quantistica, in quanto essa si differenzia dalla meccanica classica sono i seguenti:

a) non esistono in natura leggi che esprimano una successione fatale di fenomeni; anche le leggi ultime che riguardano i fenomeni elementari (sistemi atomici) hanno carattere statistico, permettendo di stabilire soltanto la *probabilità* che una misura eseguita su un sistema fisico dia un certo risultato, e ciò qualunque siano i mezzi di cui disponiamo per determinare con la maggior esattezza possibile lo stato iniziale del sistema. Queste leggi statistiche indicano un reale difetto di determinismo, e non hanno nulla di comune con le leggi statistiche classiche... Un esempio ben noto di questo nuovo tipo di leggi naturali è dato da quelle che regolano i processi radioattivi...;

b) una certa mancanza di *oggettività* nella descrizione dei fenomeni. Qualunque esperienza eseguita in un sistema atomico esercita su di esso una perturbazione finita che non può essere, per ragioni di principio, eliminata o ridotta. Il risultato di qualunque misura sembra perciò riguardare piuttosto lo stato in cui il sistema viene portato nel corso dell'esperimento stesso, che non quello inconoscibile in cui si trovava prima di essere perturbato.

La meccanica quantistica ci ha insegnato, come si diceva, a riconoscere che le trasformazioni radioattive sono guidate da una legge elementare non riducibile ad un più semplice meccanismo causale. L'introduzione nella fisica di tale nuovo tipo di leggi probabilistiche, che si nasconde, in luogo del supposto determinismo, sotto le leggi statistiche ordinarie, ci obbliga a rivedere l'analogia che abbiamo più sopra indicato con le leggi statistiche sociali.

È indiscutibile che il carattere statistico di queste ultime deriva almeno in parte dalla maniera in cui vengono definite le condizioni dei fenomeni: maniera generica, cioè propriamente "statistica". D'altra parte, se ricordiamo quanto si è detto più sopra sulle *tavole di mortalità* degli atomi radioattivi, siamo indotti a chiederci se non esista anche qui un'analogia reale con i fatti sociali, che si descrivono con linguaggio alquanto simile.

Bastano comuni artifici di laboratorio per preparare una catena comunque complessa e vistosa di fenomeni che sia *comandata* dalla disintegrazione accidentale di un solo atomo radioattivo. *Non vi è nulla dal punto di vista strettamente scientifico che impedisca di considerare come plausibile che all'origine di avvenimenti umani possa trovarsi un fatto vitale egualmente semplice, invisibile e imprevedibile*. Se è così, *come noi riteniamo*, le leggi statistiche delle scienze sociali vedono accresciuto il loro ufficio, che non è soltanto quello di stabilire empiricamente la risultante di un gran numero di cause sconosciute, ma sopratutto di dare della realtà una testimonianza immediata e concreta. La cui interpretazione richiede un'arte speciale, non ultimo sussidio dell'arte di governo.

Ettore Majorana

*(riduzione di Erasmo Recami)*

*N.B.: Segue una Seconda Parte con la trascrizione integrale dell'articolo:*

# PARTE SECONDA

(TRASCRIZIONE INTEGRALE DELL'ARTICOLO)

**In questa Seconda Parte pubblichiamo la versione integrale dell'articolo di Ettore Majorana. Per la Presentazione e il Riassunto, si veda quanto premesso (in italiano, e in inglese) agli inizi del presente scritto.** Dato che il testo apparso su *Scientia* conteneva alcuni evidenti errori, commessi nell'interpretazione della calligrafia di Majorana, anche la presente versione integrale è stata leggerissimamente "corretta":

## ETTORE MAJORANA:

### Il valore delle Leggi Statistiche nella Fisica

### e nelle Scienze Sociali

*Riassunto dell'Autore*: La concezione *deterministica* della natura racchiude in sè una reale causa di debolezza nell'irrimediabile contraddizione che essa incontra con i dati più certi della nostra stessa coscienza. G. Sorel tentò di comporre questo dissidio con la distinzione tra *natura artificiale* e *natura naturale* (quest'ultima acausale), ma negò cosi l'unità della scienza. D'altra parte l'analogia formale tra le leggi statistiche della Fisica e quelle delle Scienze Sociali accreditò l'opinione che anche i fatti umani sottostassero a un rigido determinismo. E' importante, quindi, che i recenti principii della Meccanica Quantistica abbiano portato a riconoscere (oltre ad una certa assenza di oggettività nella descrizione dei fenomeni) il carattere *statistico* anche delle leggi ultime dei processi elementari. Questa conclusione ha reso sostanziale l'analogia tra fisica e scienze sociali, tra le quali è risultata un'identità di valore e di metodo.

**********

Lo studio dei rapporti, veri o supposti, che passano fra la fisica e le altre scienze, ha sempre rivestito un notevole interesse in ragione dell'influenza speciale che la fisica ha esercitato nei tempi moderni sul generale indirizzo del pensiero scientifico. È noto che le leggi della meccanica, in modo particolare, sono apparse lungamente come il tipo insuperabile delle nostre conoscenze della natura, e si è anzi creduto da molti che a tal tipo, in ultima analisi, si sarebbero dovute ricondurre anche le nozioni imperfette fornite dalle altre scienze. Valga ciò di giustificazione allo studio che intraprendiamo.

## 1 - LA CONCEZIONE DELLA NATURA SECONDO LA FISICA CLASSICA.

Il credito eccezionale goduto dalla fisica deriva evidentemente dalla scoperta delle così dette leggi esatte, consistenti in formule relativamente semplici che, escogitate originariamente in base a indicazioni frammentarie e approssimative dell'esperienza, si rivelano in seguito di universale validità, sia che vengano applicate a nuovi ordini di fenomeni, sia che il progressivo affinamento dell'arte sperimentale le sottoponga a un controllo sempre più rigoroso. È a tutti noto che secondo la concezione fondamentale della meccanica classica, il movimento di un corpo materiale è *interamente determinato* dalle condizioni iniziali (posizione e velocità) in cui il corpo si trova, e dalle forze che agiscono su di esso. Sulla natura e misura delle forze che si possono creare nei sistemi materiali, le leggi generali della meccanica stabiliscono però in modo naturale solo qualche condizione, o limitazione, che deve essere sempre soddisfatta. Tale carattere è posseduto per esempio dal principio dell'uguaglianza fra l'azione e la reazione, al quale si sono aggiunte, in epoca meno remota, altre regole generali, come quelle riguardanti i sistemi vincolati (principio dei lavori virtuali) o le reazioni elastiche, e ancora più recentemente, con l'interpretazione meccanica del calore, anche il principio della conservazione dell'energia in quanto principio generale della meccanica. A parte tali indicazioni generali, è però compito della fisica speciale lo scoprire volta per volta quanto occorre per l'uso effettivo dei principii della dinamica, cioè la conoscenza di tutte le forze in gioco.

In un caso tuttavia è stato possibile trovare l'espressione generale delle forze che nascono fra i corpi materiali: nel caso cioè che questi siano isolati e agiscano quindi reciprocamente solo *a distanza.* In questo caso, a prescindere dalle forze elettromagnetiche scoperte posteriormente e che si manifestano però solo in particolari condizioni, l'unica forza agente si riduce alla gravitazione universale, la cui nozione venne suggerita a Newton dall'analisi matematica delle leggi di Keplero. La legge di Newton è tipicamente applicabile allo studio dei movimenti degli astri che, essendo separati da immensi spazi vuoti, possono effettivamente influenzarsi a vicenda sole per un'apparente azione a distanza. Come è noto, tale legge è realmente sufficiente per prevedere in ogni aspetto e con esattezza meravigliosa tutto il complesso svolgimento del nostro sistema planetario. Una sola minuta eccezione, riguardante lo spostamento secolare che subisce il perielio di Mercurio, costituisce una delle maggiori prove sperimentali della recente teoria della relatività generale.

Il successo sensazionale della meccanica applicata all'astronomia ha incoraggiato naturalmente la supposizione che anche i fenomeni più complicati dell'esperienza comune debbano infine ricondursi a un meccanismo simile e solo alquanto più generale della legge di. gravitazione. Secondo tale modo di vedere, che ha dato luogo alla *concezione meccanicistica* della natura, tutto l'universo materiale si svolge obbedendo a una legge inflessibile, in modo che il suo stato in un certo istante è interamente determinato dallo stato in cui si trovava nell'istante precedente; segno che tutto il futuro è implicito nel presente, nel senso che può essere previsto con assoluta certezza purché lo stato attuale dell'universo sia interamente noto. Tale concezione pienamente deterministica della natura ha avuto in seguito numerose conferme; gli sviluppi ulteriori della fisica, dalla scoperta delle leggi dell'elettromagnetismo fino alla teoria della Relatività, hanno suggerito infatti un progressivo allargamento dei principii della meccanica classica, ma hanno, d'altra parte, vigorosamente confermato il punto essenziale, cioè la completa *causalità* fisica. Non è contestabile che si debba al determinismo il merito principale e quasi esclusivo di aver reso possibile il grandioso sviluppo moderno della scienza, anche in campi lontanissimi dalla fisica. Eppure il determinismo, che non lascia alcun posto alla libertà umana e obbliga a considerare come illusori, nel loro apparente *finalismo,* tutti i fenomeni della vita, racchiude una reale causa di debolezza: la contraddizione immediata e irrimediabile con i dati più certi della nostra coscienza.

Come il suo effettivo e, secondo ogni verosimiglianza, definitivo superamento sia avvenuto proprio nella fisica in questi ultimi anni, diremo solo più avanti; sarà anzi nostro scopo ultimo l'illustrare il rinnovamento che il concetto tradizionale delle leggi statistiche deve subire in conseguenza del nuovo indirizzo seguito dalla fisica contemporanea. Ma per il momento vogliamo ancora attenerci alla concezione classica della fisica; non solo per il suo enorme interesse storico, ma anche perché essa è ancora la sola largamente conosciuta oltre la cerchia degli specialisti.

Prima di chiudere questa parte introduttiva, crediamo opportuno ricordare che le critiche al determinismo si sono moltiplicate, sopratutto in tempi a noi abbastanza vicini. La reazione filosofica, quando è stata felice, non è uscita dal suo campo, lasciando sostanzialmente intatto, se pur circoscritto nella sua importanza, il problema propriamente scientifico. Un tentativo di risolvere quest'ultimo troviamo invece in G.Sorel [*De l'Utilité du Pragmatisme*, Cap.IV, Parigi, 1921], che rappresenta la corrente pragmatistica o pluralistica. Secondo i partigiani di questo movimento, una effettiva eterogeneità dei fenomeni natu-rali esclude che se ne possa avere una conoscenza unitaria. Ogni principio scientifico sarebbe quindi applicabile a un determinato ambito di fenomeni, senza poter mai aspirare ad una validità universale. G.Sorel svolge in modo particolare la critica del determinismo, affermando che questo riguarderebbe soltanto i fenomeni che egli chiama della *natura artificiale,* caratterizzati dal fatto che essi *non* sono accompagnati da una apprezzabile *degradazione* di energia (nel senso del secondo principio della termodinamica). Tali fenomeni hanno luogo talvolta spontaneamente in natura, specie nel campo astronomico, e costituiscono allora materia di semplice osservazione; ma più frequentemente vengono provocati nei laboratori dagli sperimentatori, i quali pongono una cura particolare nell'eliminazione delle resistenze passive. Gli altri fenomeni, quelli cioè dell'esperienza comune o della *natura naturale,* nei quali entrano in gioco le resistenze passive, non sarebbero dominati da leggi definite, ma dipenderebbero in misura più o meno ampia dal caso. Il Sorel si richiama esplicitamente ad un principio metafisico di G.B.Vico. Non vogliamo qui discutere l'accentuazione arbitraria data a un particolare aspetto della scienza quale si presentava in un'epoca che non è più la nostra; dobbiamo invece rilevare che il principio pragmatista, di giudicare le dottrine scientifiche in base alla loro reale utilità, non giustifica in alcun modo la pretesa di condannare l'ideale dell'unità della scienza, che si è rivelata più volte un efficace stimolo al progresso delle idee.

## 2 - IL SIGNIFICATO CLASSICO DELLE LEGGI STATISTICHE E LE STATISTICHE SOCIALI.

Per bene intendere il significato delle leggi statistiche secondo la Meccanica, bisogna richiamarsi ad una ipotesi sulla struttura della materia che, già familiare agli antichi, entrò effettivamente nel dominio della scienza ai primi del secolo scorso per opera di Dalton; questi riconobbe per primo in tale ipotesi la naturale spiegazione delle leggi generali della chimica, da poco messe in luce. Secondo la moderna teoria atomica, che è stata definitivamente confermata con i metodi propri della fisica, esistono in natura tante specie di particelle elementari indivisibili, o *atomi,* quanti sono i corpi chimici semplici; dall'unione di due o più atomi di specie uguale o diversa, o talvolta da atomi isolati, risultano le *molecole,* le quali sono le ultime particelle capaci di una esistenza indipendente in cui si può suddividere una sostanza chimicamente definita. Le singole molecole (e talvolta anche gli atomi all'interno delle molecole), lungi dall'occupare una posizione fissa, sono animate da un movimento rapidissimo di traslazione e di rotazione su se stesse. La struttura molecolare dei corpi gassosi è particolarmente semplice. Infatti nei gas in condizioni ordinarie le singole molecole si possono considerare come particolarmente indipendenti, e a distanze reciproche considerevoli rispetto alle loro ridottissime dimensioni; segue, per il principio di inerzia, che il loro moto di traslazione è rettilineo e· uniforme, subendo modificazioni quasi istantanee nella direzione e nella misura della velocità solo in occasione di urti reciproci. Se supponiamo di conoscere esattamente le leggi che regolano l'influenza mutua delle molecole, dobbiamo attenderci, secondo i principii generali della meccanica, che basti *inoltre* conoscere nell'istante iniziale la disposizione di tutte le molecole e le loro velocità di traslazione e di rotazione, per poter prevedere *in principio* (se anche, cioè, a mezzo di calcoli troppo complessi per venire praticamente realizzati) quali saranno le esatte condizioni del sistema dopo un certo tempo. L'uso dello schema deterministico proprio della meccanica subisce tuttavia una reale limitazione di principio quando teniamo conto che i metodi *ordinari* di osservazione non sono in grado di farci conoscere esattamente le condizioni istantanee del sistema, ma ci danno solo un certo numero di informazioni globali. Dato, ad esempio, il sistema fisico risultante da una certa quantità di un determinato gas, basta conoscerne la pressione e la densità perchè risultino determinate tutte quelle altre grandezze, come temperatura, coefficiente di viscosità, ecc., che potrebbero essere oggetto di particolari misure. In altri termini, il valore della pressione e della densità bastano in questo caso a determinare interamente lo stato del sistema *dal punto di vista macroscopico,* pur non essendo evidentemente sufficienti a stabilire in ogni istante la sua esatta struttura interna, cioè la distribuzione delle posizioni e velocità di tutte le sue molecole.

Per esporre con chiarezza e brevità, e senza alcun apparato matematico, la natura del rapporto che passa fra *stato macroscopico* (A) e stato reale *(a)* di un sistema, e per trarne alcune deduzioni, dobbiamo sacrificare alquanto la precisione, pur evitando di alterare in modo essenziale la vera sostanza dei fatti. Dobbiamo dunque intendere che allo stato macroscopico A corrisponda un gran numero di possibilità effettive *a, a', a''*.... tra le quali le nostre osservazioni non ci permettono di distinguere. Il *numero* N di queste possibilità interne, secondo le concezioni propriamente classiche sarebbe ovviamente infinito, ma la teoria dei quanti ha introdotto nella descrizione dei fenomeni naturali un'essenziale discontinuità in virtù della quale il numero (N) di tali possibilità nella struttura intima di un sistema materiale è realmente *finito,* sebbene naturalmente grandissimo. Il valore di N dà una misura del grado di indeterminazione *nascosta* del sistema; è però praticamente preferibile considerare una grandezza proporzionale al suo logaritmo, ovvero $S = k \log N$, essendo k la costante universale di Boltzmann, determinata in modo che S coincida con una grandezza fondamentale, già nota, della termodinamica: *l'entropia.* L'entropia si presenta in realtà come una grandezza fisica al pari del peso, dell'energia ecc., sopra tutto perché come quest'altre grandezze gode della proprietà additiva: cioè la entropia di un sistema risultante da più parti indipendenti è uguale alla somma delle entropie delle singole parti. Per dimostrarlo, basta, da un lato, osservare che il numero di possibilità latenti di un sistema composto è uguale evidentemente al prodotto dei numeri analoghi relativi alle parti costituenti; e tener presente, dall'altro, la nota regola elementare che stabilisce la corrispondenza fra il prodotto di due o più numeri e la somma dei rispettivi logaritmi.

Sul modo di determinare il complesso di configurazioni interne *a, ·a', a''* ..... che corrisponde allo stato macroscopico A, non sorgono in genere difficoltà. Si può invece discutere se tutte le singole possibilità *a, a', a''* .... si debbano o no riguardare come egualmente probabili. Orbene, secondo un'ipotesi che si ha ragione di credere generalmente verificata ( detta ipotesi ergodica, o quasi ergodica), se un sistema persiste *indefinitamente* in uno stato A, allora si può affermare che esso passa un'eguale frazione del suo tempo in ciascuna delle configurazioni *a, a', a''*....; si è così condotti a considerare effettivamente come egualmente probabili tutte le possibili determinazioni interne. È questa in realtà una nuova ipotesi, poiché l'universo, lungi dal permanere indefinitamente nello stesso stato, va soggetto a trasformazioni continue. Ammetteremo dunque, come ipotesi di lavoro estremamente plausibile, ma le cui conseguenze lontane potrebbero anche talvolta non essere verificate, che tutti i possibili stati interni di un sistema in condizioni fisiche determinate siano a priori egualmente probabili. Risulta così interamente definito il *complesso statistico* associato ad ogni stato macroscopico A.

Il problema generale della meccanica statistica si può così riassumere: essendo definito statisticamente, come si è detto, lo stato A iniziale del sistema, quali previsioni sono possibili in riguardo al suo stato al tempo *t* ? Può apparire a prima vista che questa definizione sia troppo ristretta, poiché oltre al problema propriamente dinamico altri se ne possono considerare di carattere *statico;* ad es. qual è la temperatura di un gas di cui siano noti le pressioni e la densità? E così in tutti i casi in cui si voglia, da alcune caratteristiche di un sistema, sufficienti a definirne lo stato, dedurne altre che possano interessare. La distinzione si può peraltro formalmente ignorare: incorporando infatti nel sistema appropriati strumenti di misura, ci si può sempre ricondurre al caso precedente.

Supponiamo dunque che lo stato iniziale del sistema in esame risulti da un complesso statistico A == *(a, a', a''* ..... *)* di casi possibili e, per quanto si è detto, egualmente probabili.

Ciascuna di queste determinazioni concrete si modifica nel corso del tempo secondo una legge che, in accordo con i principii generali della meccanica, dobbiamo ancora ritenere rigidamente causale, cosicché dopo un certo tempo si passa dalla serie *a, a', a''* ... a un'altra serie ben determinata *b,b',b''*....; il complesso statistico *(b, b', b''* ..... *),* che è anch'esso costituito da N elementi egualmente probabili come il complesso originario A (teorema di Liouville), definisce tutte le possibili previsioni sullo svolgimento del sistema. Per ragioni che solo un'analisi matematica complessa potrebbe precisare, accade in generale che tutti i casi semplici appartenenti alla serie $b, b', b''...$ *salvo un numero del tutto insignificante di eccezioni*, costituiscono *in tutto* o *in parte* un nuovo complesso statistico B definito come A da uno stato *macroscopicamente* ben determinato. Possiamo allora enunciare la *legge statistica* secondo la quale vi è la pratica certezza che il sistema debba passare da A in B. Per quanto si è detto, il complesso statistico B è almeno così ampio come A, cioè contiene un numero di elementi non inferiore a N; segue che l'entropia di B è uguale a quella di A *o maggiore.* Durante qualunque trasformazione che si compia spontaneamente *in accordo con le leggi statistiche* si ha quindi costanza o aumento di entropia, mai diminuzione: è questo il fondamento statistico del famoso secondo principio della termodinamica.

È notevole che dal punto di vista pratico il passaggio da A a B si può considerare come certo; ciò che spiega come storicamente le leggi statistiche siano state considerate dapprima altrettanto fatali delle leggi della meccanica e solo per il progresso dell'indagine teorica se ne sia in seguito riconosciuto il vero carattere. Le leggi statistiche abbracciano gran parte della fisica. Fra le applicazioni più note ricordiamo: l'equazione di stato dei gas, la teoria della diffusione, della conducibilità termica, della viscosità, della pressione osmotica e molte altre consimili. Un posto a parte merita la teoria statistica dell'irraggiamento che introduce per la prima volta nella fisica il *discontinuo* simboleggiato della costante di Planck. Ma vi è inoltre una intera branca della fisica, la *termodinamica*, i cui principii, benché fondati direttamente sull'esperienza, si possono ricondurre alle nozioni generali della meccanica statistica. Per quanto abbiamo fatto finora, si può così riassumere il significato delle leggi statistiche secondo la fisica classica: 1°) i fenomeni naturali obbediscono ad un determinismo assoluto; 2°) l'osservazione *ordinaria* non permette di riconoscere esattamente lo stato interno di un corpo, ma solo di stabilire un complesso innumerevole di possibilità indistinguibili; 3°) stabilite delle ipotesi plausibili sulla probabilità delle diverse possibilità, e supposte valide le leggi della meccanica, il calcolo delle probabilità permette la previsione più o meno certa dei fenomeni futuri. Possiamo ormai esaminare il rapporto che passa fra le leggi stabilite dalla meccanica classica e quelle regolarità francamente empiriche che sono note con lo stesso nome in modo particolare nelle scienze sociali.

Bisogna anzitutto convincersi che l'analogia formale non potrebbe essere più stretta. Quando si enuncia, ad es., la legge statistica: «In una società moderna di tipo europeo il coefficiente annuo di nuzialità è prossimo a 8 per 1000 abitanti», è abbastanza chiaro che il sistema su cui dobbiamo eseguire le nostre osservazioni è definito solo in base a certi caratteri globali rinunciando deliberatamente a indagare tutti _quei dati ulteriori

(come per es. la biografia di tutti gli individui che compongono la società in esame) la cui conoscenza sarebbe indubbiamente utile per prevedere il fenomeno con maggiore precisione e sicurezza di quanto non consenta la generica legge statistica; non altrimenti, allorché si definisce lo stato di un gas semplicemente dalla pressione e dal volume, si rinunzia deliberatamente a investigare le condizioni iniziali di tutte le singole molecole. Una differenza sostanziale si potrebbe invece scorgere nel carattere matematicamente definito dalle leggi statistiche della fisica a cui fa riscontro quello chiaramente empirico delle leggi statistiche sociali; ma è plausibile attribuire l'empirismo delle statistiche sociali (intendiamo precisamente l'incostanza dei loro risultati *oltre la parte spettante al caso*) alla complessità dei fenomeni che esse considerano, per cui non è possibile definire esattamente le condizioni o il contenuto della legge. D'altra parte anche la fisica conosce le leggi empiriche quando studia fenomeni di puro interesse applicativo; tali, ad es., le leggi sull'attrito fra corpi solidi, o sulle proprietà magnetiche dei vari tipi di ferro e altri simili. Infine si potrebbe dare speciale importanza alla differenza nei metodi di rilevazione, che nella fisica sono globali (così basta lettura di uno strumento di misura per conoscere la pressione di un gas benché essa derivi dalla somma degli impulsi indipendenti che le singole molecole trasmettono alle pareti), mentre nelle statistiche sociali si registrano di solito i fatti individuali; non è però neanche questa un'antitesi assoluta, come prova la possibilità dei metodi più vari di rilevazione indiretta. Ammesse così le ragioni che fanno credere all'esistenza di una reale analogia fra le leggi statistiche fisiche e sociali, siamo indotti a ritenere plausibile che, come le prime presuppongono logicamente un rigido determinismo, così le ultime siano da parte loro la prova più diretta che il più assoluto determinismo governa anche i fatti umani; argomento che ha avuto tanto miglior fortuna in quanto, come abbiamo detto in principio, si era manifestata per ragioni indipendenti la tendenza a vedere nella causalità della fisica classica un modello di valore universale. Sarebbe qui fuor di luogo riprendere discussioni antiche e mai concluse, ma crediamo di poter ricordare, come fatto generalmente ammesso, che la non avvenuta conciliazione fra le nostre contrastanti intuizioni della natura ha lungamente pesato sul pensiero moderno e sui valori morali. Non va quindi accolto semplicemente come una curiosità scientifica l'annunzio che negli ultimissimi anni la fisica é stata costretta ad abbandonare il suo indirizzo tradizionale rigettando, in maniera verosimilmente definitiva, il determinismo assoluto della meccanica classica.

## 3. - LE NUOVE CONCEZIONI DELLA FISICA.

È impossibile esporre con qualche compiutezza in poche righe lo schema matematico e il contenuto sperimentale della meccanica quantistica [il lettore che desideri approfondire le sue conoscenze in tale materia aggirando, finché si può, lo scoglio matematico, può consultare W.Heisenberg, *Die Physikalischen Prinzipien del Quantentheorie,* Lipsia 1930]. Ci limiteremo pertanto a qualche accenno. Vi sono dei fatti sperimentali noti da gran tempo (fenomeni di interferenza) che depongono irrefutabilmente a favore della teoria ondulatoria della luce; altri fatti scoperti da recente (effetto Compton) suggeriscono, al contrario, non meno decisivamente l'opposta teoria corpuscolare. Tutti i tentativi di comporre la contraddizione nel quadro della fisica classica sono rimasti assolutamente infruttuosi: il che può anche sembrare poco significativo. Senonché di tali fatti inesplicabili, e di altri non meno inesplicabili e della più diversa natura, e infine di *quasi tutti* i fenomeni noti ai fisici e finora insufficientemente spiegati, si è trovata realmente da pochi anni la spiegazione unica e meravigliosamente semplice: quella contenuta nei principii della meccanica quantistica. Questa straordinaria teoria è dunque così solidamente fondata nell'esperienza come forse nessun'altra fu mai; le critiche a cui essa fu ed è assoggettata non possono quindi concernere in alcun modo la legittimità del suo uso per l'effettiva previsione dei fenomeni, ma soltanto l'opinione, condivisa dai più, che il nuovo indirizzo da essa segnato debba conservarsi, e anzi ancora accentuarsi, nei futuri sviluppi della fisica. Gli aspetti caratteristici della meccanica quantistica, in quanto essa si differenzia dalla meccanica classica sono i seguenti:

*a)* non esistono in natura leggi che esprimano una successione fatale di fenomeni; anche le leggi ultime che riguardano i fenomeni elementari (sistemi atomici) hanno carattere statistico, permettendo di stabilire soltanto la *probabilità* che una misura eseguita su un sistema preparato in un dato modo dia un certo risultato, e ciò qualunque siano i mezzi di cui disponiamo per determinare con la maggior esattezza possibile lo stato iniziale del sistema. Queste leggi statistiche indicano un reale difetto di determinismo, e non hanno nulla di comune con le leggi statistiche classiche nelle quali l'incertezza dei risultati deriva dalla volontaria rinunzia, per ragioni pratiche, a indagare nei più minuti particolari le condizioni iniziali dei sistemi fisici. Vedremo più avanti un esempio ben noto di questo nuovo tipo di leggi naturali;

*b)* una certa mancanza di *oggettività* nella descrizione dei fenomeni. Qualunque esperienza eseguita in un sistema atomico esercita su di esso una perturbazione finita che non può essere, per ragioni di principio, eliminata o ridotta. Il risultato di qualunque misura sembra perciò riguardare piuttosto lo stato in cui il sistema viene portato nel corso dell'esperimento stesso che non quello inconoscibile in cui si trovava prima di essere perturbato. Questo aspetto della meccanica quantistica è senza dubbio più inquietante, cioè più lontano dalle nostre intuizioni ordinarie, che non la semplice mancanza di determinismo.

Fra le leggi probabilistiche riguardanti i fenomeni elementari è nota da più antica data quella che regola i processi radioattivi.

Ogni atomo di una sostanza radioattiva ha una probabilità definita *m*d*t* di trasformarsi nel tempuscolo d*t* in seguito all'emissione, o di una particella *alfa* (nucleo di elio), ovvero in altri casi di una particella *beta* (elettrone). *Il tasso di mortalità, m,* è costante, cioè indipendente dalla *età* dell'atomo, ciò che dà una forma particolare (esponenziale) alla *curva di sopravvivenza;* la vita media vale 1/*m* e in modo elementare si può determinare analogamente la *vita probabile*, chiamata talvolta *periodo di trasformazione.* Entrambe sono indipendenti dall'età dell'atomo che non manifesta del resto per alcun altro segno un reale invecchiamento con il progredire del tempo. Esistono vari metodi per l'osservazione, o anche per la registrazione automatica delle singole trasformazioni che avvengono nel seno di una sostanza radioattiva, ed è stato quindi possibile verificare, mediante dirette rilevazioni statistiche e applicazioni del calcolo della probabilità, che i singoli atomi radioattivi non subiscono alcuna influenza reciproca o esterna per quanto riguarda l'istante della trasformazione; infatti il numero delle disintegrazioni che hanno luogo in un certo intervallo di tempo è soggetto a fluttuazioni dipendenti esclusivamente dal caso, cioè dal carattere probabilistico della legge individuale di trasformazione.

La meccanica. quantistica ci ha insegnato a vedere nella legge esponenziale delle trasformazioni radioattive una legge elementare non riducibile .ad un più semplice meccanismo causale. Naturalmente anche le leggi statistiche note alla meccanica classica e riguardanti *sistemi complessi,* conservano la loro validità secondo la meccanica quantistica. Quest'ultima modifica peraltro le regole per la determinazione delle configurazioni interne, e in due modi diversi, a seconda della natura dei sistemi fisici, dando luogo rispettivamente alle teorie statistiche di Bose-Einstein, o di Fermi. Ma l'introduzione nella fisica di un nuovo tipo di legge statistica, o meglio semplicemente probabilistica, che si nasconde, in luogo del supposto determinismo, sotto le leggi statistiche ordinarie, obbliga a rivedere le basi dell'analogia che abbiamo stabilita più sopra con le leggi statistiche sociali. È indiscutibile che il carattere statistico di queste ultime deriva almeno in parte dalla maniera in cui vengono definite le condizioni dei fenomeni: maniera generica, cioè propriamente statistica., e tale da permettere un complesso innumerevole di possibilità concrete differenti. D'altra parte, se ricordiamo quanto si è detto più sopra sulle *tavole di mortalità* degli atomi radioattivi, siamo indotti a chiederci se non esista anche qui un'analogia reale con i fatti sociali, che si descrivono con linguaggio alquanto simile.

Qualche cosa a prima vista sembra escluderlo; la disintegrazione di un atomo è un fatto semplice, imprevedibile, che avviene improvvisamente e isolatamente dopo un'attesa talvolta di migliaia e perfino di miliardi di anni; mentre niente di simile accade per i fatti registrati dalle statistiche sociali. Questa non è però un'obiezione insormontabile.

La disintegrazione di un. atomo radioattivo può obbligare un contatore automatico a registrarlo con effetto meccanico, reso possibile da adatta amplificazione. Bastano quindi comuni artifici di laboratorio per preparare una catena comunque complessa e vistosa di fenomeni che sia *comandata* dalla disintegrazione accidentale di un solo atomo radioattivo. Non vi è nulla dal punto·di vista strettamente scientifico che impedisca di considerare come plausibile che all'origine di avvenimenti umani possa trovarsi un fatto vitale egualmente semplice, invisible e imprevedibile. Se è così, come noi riteniamo, le leggi statistiche delle scienze sociali vedono accresciuto il loro ufficio, che non è soltanto quello di stabilire empiricamente la risultante di un gran numero di cause sconosciute, ma sopratutto di dare della realtà una testimonianza immediata e concreta. La cui interpretazione richiede un'arte speciale, non ultimo sussidio dell'arte di governo.

(Ettore Majorana)

*********************************************

## BIBLIOGRAFIA:

L'articolo di Majorana (a suo tempo da lui cestinato) ha visto la luce solo postumo, nel 1942 (Ettore Majorana scomparve misteriosamente, come noto, il 26 marzo 1938), per desiderio dei fratelli di Ettore e per interessamento di Giovanni Gentile jr.:

[2] E.Majorana: "Il valore delle Leggi Statistiche nella Fisica e nelle Scienze Sociali" (a cura di G.Gentile jr.), *Scientia,* vol.36, fascicolo del Febbraio-Marzo del 1942, pp.58-66.

Successivamente non è stato più ripubblicato, in italiano, fino agli inizi del 2006 (centenario della nascita del Nostro), quando ne abbiamo reso note alcune riduzioni, apparsi su quotidiani italiani (di Catania, Bergamo, Torino, etc.) e sulla rivista "Fisica in Medicina"; si vedano, a questo proposito, e per esempio,

[3] E.Recami (a cura, e con presentazione di): "E.Majorana: Il valore delle scienze statistiche nella fisica e nelle scienze sociali", *La Sicilia* (Catania; 1 marzo 2006), pagina culturale (p.21),

e, più ancora:

[4] E.Recami: "Nel centenario della nascita di Ettore Majorana": presentazione, e riduzione, a cura di E.Recami dell'articolo di E.Majorana, Il valore delle Scienze Statistiche nella Fisica e nelle Scienze Sociali", *Fisica in Medicina* (anno 2006), fascicolo n.3, pp.261-265.

Per le traduzioni (delle quali forse ne esistono anche una in francese e un'altra in inglese) si vedano

[5] "The tenth article of Ettore Majorana", presentazione di R.N.Mantegna di una traduzione in inglese organizzata da J.Bettany; il tutto pubblicato su *Quantitative Finance*, vol.5 (2005) p.133; e

[6] R.N.Mantegna: "The value of statistical laws in physics and in social sciences", Cap.10 in *Ettore Majorana – Scientific Papers,* a cura di G.F.Bassani et al. (S.I.F. e Sprinter; Bologna e Berlino, 2006), p.250.

Tutti i documenti noti circa E.Majorana –tranne quelli rinvenuti negli ultimi due anni—sono stati scoperti e raccolti, e per primo pubblicati, da E.Recami: inclusa la lettera sopra citata del Nostro all'amico G.Gentile jr; si veda, ad es., il libro uscito nella sua prima edizione nel 1987

[7] E.Recami: *Il Caso Majorana: Epistolario, Testimonianze, Documenti,* seconda edizione ("Oscar" Mondatori, Mondatori; Milano 1991), di 230 pagine; e, in particolare, quarta edizione (Di Renzo Editore; Roma, 2002), di 273 pagine. [Il materiale ivi contenuto è protetto fin dal 1986 da copyright a favore di Maria Majorana, in solido con E.Recami e, ora, con la Di Renzo Editore]. Della seconda edizione esiste una pregevole

traduzione in francese ad opera di F. e Ph. Gueret (inedita).

Il citato volume [7] contiene anche una panoramica sul lavoro scientifico; per informazioni biografiche, e ancor più scientifiche, si vedano anche [oltre alla homepage www.unibg.it/recami] gli e-prints dello stesso autore

[8] E.Recami et al., arXiv: physics/9810023v4 [physics.hist-ph]; arXiv: 0708.2855v1 [physics.hist-ph]; e arXiv: 0709.1183 [physics.hist-ph].

Per quanto riguarda la scoperta delle *parastatistiche* da parte di Giovanni Gentile jr. (da lui chiamate "statistiche intermedie"), si vedano gli articoli

[9] G.Gentile jr., "Osservazioni sopra le statistiche intermedie", *Rendiconti Istituto Lombardo,* vol.74 (1940-41) pp.133-137; oltre a *Nuovo Cimento*, vol.17 (1940) p.493; e

[10] G.Gentile jr., "Sopra il fenomeno della condensazione del gas di Bose-Einstein", *Ricerca Scientifica*, vol.12 (1941) 341-346. Purtroppo G.Gentile jr. morì prematuramente nel 1942.